\documentclass[12pt]{article}
\usepackage{epsfig}
\usepackage{amsfonts}
\usepackage{amsopn}
\usepackage{amsmath}

\newcommand{\eq}{\begin{equation}}
\newcommand{\eqx}{\end{equation}}
\newcommand{\eqn}{\begin{eqnarray}}
\newcommand{\eqnx}{\end{eqnarray}}
\newcommand{\f}[2]{\frac{#1}{#2}}

\newcommand{\lm}{\lambda}
\renewcommand{\th}{\theta}

\newcommand{\dl}{\delta}

\newcommand{\sg}{\sigma}
\newcommand{\Sg}{\Sigma}
\newcommand{\eps}{\varepsilon}
\newcommand{\qqqq}{\quad\quad\quad\quad}

\newcommand{\etat}{\tilde{\eta}}
\newcommand{\chit}{\tilde{\chi}}
\newcommand{\qt}{\tilde{q}}
\newcommand{\qs}{q_*}
\newcommand{\ps}{p_*}
\newcommand{\qts}{\tilde{q}_*}
\newcommand{\sqg}{\sqrt{8g^2}}
\newcommand{\sinpt}{\sin \f{p}{2}}
\newcommand{\ZZ}{\mathbb{Z}}
\DeclareMathOperator*{\res}{res}
\DeclareMathOperator{\arcsinh}{arcsinh}
\DeclareMathOperator{\arctanh}{arctanh}
\DeclareMathOperator{\tr}{tr}
\newcommand{\OO}[1]{{\cal O}\left(#1\right)}
\newcommand{\cor}[1]{\left\langle #1 \right\rangle}

\title{Wrapping interactions at strong coupling\\ 
-- the giant magnon} 

\author{Romuald A. Janik\thanks{e-mail: {\tt ufrjanik@if.uj.edu.pl}}
\  and Tomasz {\L}ukowski\thanks{e-mail: {\tt tomaszlukowski@gmail.com}} \\ \\
Institute of Physics\\
Jagellonian University,\\
ul. Reymonta 4, \\
30-059 Krak{\'o}w\\
Poland
}

\begin{document}

\maketitle

\begin{abstract}
We derive generalized L\"{u}scher formulas for finite size corrections
in a theory with a general dispersion relation. For the $AdS_5 \times
S^5$ superstring these formulas encode leading wrapping interaction effects.
We apply the generalized $\mu$-term formula to calculate finite size
corrections to the dispersion relation of the giant magnon at strong
coupling. The result exactly agrees with the classical string computation of
Arutyunov, Frolov and Zamaklar. The agreement involved a Borel
resummation of all even loop-orders of the BES/BHL dressing factor
thus providing a strong consistency check for the choice of the
dressing factor. 
\end{abstract}

\vfill

\pagebreak

\section{Introduction}

One of the most interesting and rapidly developing lines of investigation in
recent years has been the study of integrable structures discovered on
both sides
\cite{Minahan:2002ve,Beisert:2003tq,Beisert:2003yb,kor1,Bena:2003wd,
Dolan:2003uh}  
of the AdS/CFT correspondence \cite{Maldacena:1997re}. The recent developments
allow for an interpolation all the way from weak to strong coupling
moving us closer to the complete knowledge of the anomalous dimensions
of all operators on the ${\cal N}=4$ SYM gauge theory side, and the
energy spectrum of the quantized superstring in $AdS_5 \times S^5$. 

A lot is currently known about the properties of the integrable
worldsheet theory of the superstring in $AdS_5 \times S^5$ on the
plane. The full exact S-matrix is now believed to be known. Initially
its structure in various subsectors of the theory has been uncovered
\cite{S,BS}, which culminated with \cite{Beisert} where the $su(2|2)
\times su(2|2) \subset psu(2,2|4)$ symmetry has been exploited to
determine the S-matrix up to a scalar function
\eq
S(p_1,p_2)=S_0(p_1,p_2) \cdot \left[ \hat{S}_{su(2|2)}(p_1,p_2)
\otimes \hat{S}_{su(2|2)}(p_1,p_2) \right]
\eqx  
The function $S_0(p_1,p_2)$ is related to the the dressing factor
$\sg^2(p_1,p_2)$ which is 1 at weak coupling up to three loops and
whose leading and subleading behaviour at strong coupling has been
determined in \cite{AFS} and \cite{HL} respectively.

The task of fixing the dressing factor at all values of the coupling
has been concluded in \cite{BES}, where a specific solution of
crossing constraints \cite{CROSSING} in \cite{BHL} was chosen based on
arguments of transcendentality \cite{Lipatov,ES,BES}. A very
nontrivial cross-check of this choice was the 4-loop calculation of
\cite{Bern}.  

The BES/BHL dressing factor satisfies all known constraints both at
weak and at strong coupling, yet it is not known how unique is this
choice. So it is also interesting to independently test as large part
of this solution as possible. A 2-loop test has been performed in the
near-flat space limit in \cite{Zarembotwoloop}, while the
considerations in \cite{DoreyMaldacena} on the location of double
poles involve the full expression.
As a byproduct, the present paper
provides a stringent test sensitive to {\em all} even loop orders at
strong coupling. 

Despite our almost complete knowledge of the S matrix of the theory
and the knowledge of the energies of states with large R-charge $J$
(equivalently the anomalous dimensions of long operators) through the
asymptotic Bethe ansatz, not much is known for operators for finite
$J$. On the gauge theory side, it is known \cite{BDS} that at roughly
$J$ loop order wrapping interactions appear, which are not expected to
be captured by the asymptotic Bethe ansatz. As one increases the
coupling, and keeps $J$ fixed the problem becomes more and more
severe. Recently there appeared some explicit calculations which
showed the limitations of the asymptotic Bethe ansatz both at weak
\cite{LSRV} and at strong \cite{SN,SNZZ,AFZ} coupling. It would be very
interesting to understand these phenomenae on some general grounds.

In order to proceed one may try to take here either the gauge
theory perspective using the spin chain language or the dual string
theory point of view. It seems that in this case it is this dual
string formulation which is more fruitful.

In \cite{US} it has been suggested that wrapping interactions
correspond to finite size corrections of the integrable worldsheet
quantum field theory which arise due to virtual corrections such as a
virtual particle going around the circumference of the cylinder. It
has been argued that, as it happens in conventional relativistic
integrable field theories, these finite size corrections are uniquely
determined by the knowledge of the theory at infinite size. 

In fact, in relativistic integrable field theories, leading formulas
for mass (energy) finite size correction were derived long time ago by
L\"{u}scher \cite{LUSCHER}. The leading correction is the sum of two
terms -- the F-term
\eq
\label{e.fluscher}
\Delta m_F (L) = -m \int_{-\infty}^\infty
\f{d\th}{2\pi} e^{-m L \cosh \th} \cosh \th \sum_b
\left(S_{ab}^{ab}\left(\th+i \f{\pi}{2} \right)-1 \right)
\eqx
and the $\mu$-term
\eq
\label{e.muluscher}
\Delta m_\mu (L) =  -\f{\sqrt{3}}{2} m\sum_{b,c} M_{abc} (-i)
\,{\mbox{\rm res}}_{\th=2\pi i/3}\, S_{ab}^{ab}(\th) \cdot
e^{-\f{\sqrt{3}}{2}m L}
\eqx
quoted here for a theory with a single mass scale \cite{KM} in 1+1
dimensions. $S_{ab}^{ab}(\th)$ is the (infinite volume) $S$-matrix element, and
$M_{abc}=1$ if $c$ is a bound state of $a$ and $b$ and zero
otherwise. The F-term formula has the interpretation of a virtual
particle going around the cylinder and interacting with the physical
particle. This process is depicted graphically at the right in
fig. 1. The $\mu$-term arises due to the splitting of the particle
into a pair of virtual (but on-shell) particles which then
recombine. This process is shown on the left in fig. 1. 

\begin{figure}[t]
\centerline{\scalebox{0.5}{\rotatebox{0}{\includegraphics{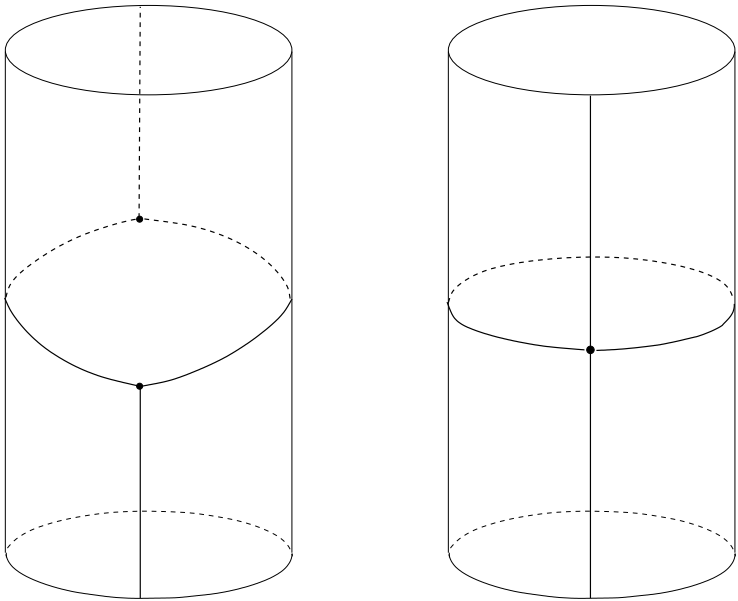}}}}
\caption[fig1]{The diagram to the left (the $\mu$-term) shows a particle
splitting in two virtual, on-shell particles, traveling around the 
cylinder and recombining. The diagram to the right (the F-term)
shows a virtual particle going around the circumference of the cylinder.}
\label{fig1}
\end{figure}

Unfortunately, we cannot use directly the above formulae since the
worldsheet theory is not relativistic. Indeed the elementary
excitations (magnons \cite{magnon}) obey the dispersion relation
\eq
\label{e.dispersion}
E=\sqrt{1+8g^2 \sin^2 \f{p}{2}}
\eqx 
where $g$ is related to the `t Hooft coupling as
$g^2=\frac{\lambda}{8\pi^2}$. In \cite{US} considerations related to
the Thermodynamic Bethe Ansatz approach to finite size corrections led
to suggestions for the form of exponential terms (magnitudes) of the
corrections for the $AdS_5 \times S^5$ worldsheet theory. However a
complete formula together with the prefactor was still missing.

The aim of this paper is to provide a diagrammatic derivation for the
leading finite size correction to the energy of a magnon and to
explicitly evaluate the $\mu$-term in this case at strong
coupling. This particular calculation is interesting since there exists
a classical string computation for the same quantity \cite{AFZ} with
which one can compare.

Finally let us note that the complexity of the dressing phase has
prompted several groups to suggest that it can be obtained in some
simpler setting by eliminating other degrees of freedom/higher levels
\cite{mannpol,kazakov,gromov,srz,sakai}. We hope that the
formalism of finite size corrections may be a strong test on the
proposed constructions as it is sensitive to all virtual particles of
the theory.

The plan of this paper is as follows. In section 2 we will review
recent results on finite size corrections both at weak and at
strong coupling. In section 3 we will show how the postulated
exponential terms reproduce the magnitudes of these corrections. In
section 4 we present a diagrammatic derivation of the finite size
corrections which generalize (\ref{e.fluscher})-(\ref{e.muluscher}) to
a theory with a quite general dispersion relation. We then apply the
generalized formula for the $\mu$-term to the case of a magnon at
strong coupling. We also consider the case of general light-cone
gauges ($a$-gauges) for the worldsheet theory. We close the paper with
conclusions and some appendices with more technical parts of the
calculations.  

\section{Finite size corrections}

As mentioned in the introduction, recently there appeared a number of
explicit computations which showed the limitations of the asymptotic
Bethe ansatz. In this section we would like to briefly review these
results.

At weak coupling, wrapping interactions appear generically at the
order $g^{2L}$, although for some operators that order may be
higher\footnote{This happens when an operator of bigger length is a
  member of the same supersymmetry multiplet.}. Thus e.g. for the
Konishi operator wrapping interactions appear at four loops. Obtaining
explicit predicitions seemed therefore to be nearly hopeless. However in
\cite{LSRV} it was shown that results from asymptotic Bethe ansatz at
four loops, i.e. exactly where we expect additional wrapping
contributions, are in conflict with perturbative expectations from BFKL.

At strong coupling, in \cite{SN,SNZZ} exponential corrections beyond
asymptotic Bethe ansatz for spinning strings in the {\sl su(2)} and
{\sl sl(2)} sector were determined. The magnitude of these corrections is
\eq
e^{-\f{2\pi J}{\sqrt{\lm}}}
\eqx  

Finally, finite size corrections were found for the giant magnon
dispersion relation at strong coupling from classical string solutions
\cite{AFZ,Gordon}. They have the form
\eq
\label{e.magstring}
\dl E_{string}=-\f{\sqrt{\lm}}{\pi} \cdot \f{4}{e^2} \cdot \sin^3 \f{p}{2}
\cdot e^{-\f{2\pi J}{\sqrt{\lm} \sinpt}} \equiv
-g \cdot \f{8\sqrt{2}}{e^2} \cdot \sin^3 \f{p}{2} \cdot
e^{-\f{1}{\sqrt{2} g \sinpt} J}
\eqx
where we gave the result both in terms of $\lm$ and in terms of $g$
which we will  
use. The curious numerical prefactor includues $e=\exp(1)$, the base of natural
logarithm. This result is even more interesting when one compares it
to finite size corrections for the magnon computed within the Hubbard
model approach \cite{hubbard} which gives \cite{hubbard,Gordon}
\eq
\label{e.maghub}
\dl E_{Hubbard}=-\f{\pi}{\sqrt{\lm}} \cdot 2 \cdot \f{1}{\sinpt} \cdot
e^{-\f{2\pi J}{\sqrt{\lm} \sinpt}} 
\eqx
We see that the exponential term is identical in both expressions,
while the prefactors differ not only in momentum dependence but also
in the scaling with $\lm$. The main motivation of the present paper is
to understand {\em quantitatively} the origin of these expressions and
to show that they follow from a worldsheet quantum field theoretical
picture of virtual corrections going around the circumference of the
(worldsheet) cylinder as advocated in \cite{US}.

\section{TBA motivated exponential terms}

In \cite{US} it was suggested that finite size corrections to
energies of string states could be found from a Thermodynamic Bethe
Ansatz reasoning following the route applied already with success to
{\em relativistic} integrable field theories initially for the ground
state \cite{TBAGROUND} and later extended to excited states
\cite{Dorey}. The TBA framework suggests a natural generalization
\cite{US} to the case of the worldsheet superstring theory which is
not relativistic due to the nonstandard dispersion relation
(\ref{e.dispersion}). Other approaches used in the relativistic
context, such as NLIE \cite{NLIE,fioravanti} or
\cite{BLZZ,Teschner} seem to be more difficult to adapt here.

The basic idea of TBA is to perform a spacetime interchange. In order
to find the energy of a (ground) state for the theory on a circle of
circumference $J$, one considers the (euclidean) partition function
\eq
Z=\tr e^{-R H}
\eqx  
with $R\to \infty$. The same quantity may be interpreted as the
partition function for the theory, with space and time interchanged, on
a very big circle of circumference $R$ at a temperature $1/J$. The
advantage is that in this case the Bethe ansatz is exact (since $R\to
\infty$) and one can use it to obtain explicit integral equations for
the exact energies. This approach was later extended to excited states
in \cite{Dorey}. 

Following these reasonings, \cite{US} suggested that the magnitude of
the exponential finite size corrections is
\eq
e^{-E_{TBA} J}
\eqx 
where $E_{TBA}=-ip$ and $p_{TBA}=-iE$. For the dispersion relation
(\ref{e.dispersion}) this leads to
\eq
\label{e.heurf}
e^{-2J \arcsinh \left(\f{1}{\sqg} \cdot \sqrt{1+p^2_{TBA}} \right)}
\eqx
This term should be understood as an analog of $e^{-L \cosh \th}$ in
the F-term (\ref{e.fluscher}). 
Generalization of the $\mu$-term is more heuristic. The $\mu$-term
comes from a particle with momentum $p$ splitting into {\em on-shell}
constituents with momenta $p_c$ and $p-p_c$. The exponential term
in\footnote{And its nonzero $p$ generalization.}
(\ref{e.muluscher}) then can be rewritten as
\eq
\label{e.heurmu}
e^{-J \cdot Im\; p_c }
\eqx
In this paper we will independently derive the complete
expressions for the generalized F- and $\mu$-terms including the
preexponential prefactors. For the moment let
us see how these heuristic expressions
(\ref{e.heurf})-(\ref{e.heurmu}) fit the results reviewed in the
previous section. Some of these results were already discussed in
\cite{US}.

At weak coupling, when $g$ is small, the $\arcsinh$ behaves like
$-\log g+\ldots$, therefore \cite{US}
\eq
e^{-L E_{TBA}} \sim g^{2L} \cdot \ldots
\eqx
exactly as expected for wrapping interactions.

At strong coupling, the argument of $\arcsinh$ is small so effectively
we get \cite{US}
\eq
e^{-L E_{TBA}} \sim e^{-\f{L}{\sqrt{2} g} \sqrt{1+p^2_{TBA}}} \sim
e^{-\f{L}{\sqrt{2} g}} \equiv e^{-\f{2\pi L}{\sqrt{\lm}}}
\eqx
which is just the exponential term in the correction for the spinning
string in \cite{SN,SNZZ}.

Finally let us consider (\ref{e.heurmu}) at strong coupling. We have
to solve the on-shell condition\footnote{In fact the particle with
  momentum $p_c$ could be the BPS bound state with energy
  $\sqrt{4+\f{\lm}{\pi^2} \sin^2 \f{p_c}{2}}$. This does not change
  the subsequent results as long as the particle with momentum $p-p_c$
  is just a magnon.}  
\eq
\sqrt{1+\f{\lm}{\pi^2} \sin^2 \f{p}{2}} =
\sqrt{1+\f{\lm}{\pi^2} \sin^2 \f{p_c}{2}} +
\sqrt{1+\f{\lm}{\pi^2} \sin^2 \f{p-p_c}{2}}
\eqx
Perturbatively we find
\eq
p_c=p+\f{2\pi i}{\sqrt{\lm} \sinpt}
\eqx
Plugging this back into (\ref{e.heurmu}) we obtain
\eq
\label{e.expmu}
e^{-J \cdot Im\; p_c }=e^{-\f{2 \pi J}{\sqrt{\lm} \sinpt}}
\eqx
which is exactly the exponential term appearing in the finite size
corrections to the giant magnon dispersion relation. From the above
derivation we see that this term is very generic and depends only on
the dispersion relation (\ref{e.dispersion}). Thus it is not
surprising that both the 
string result (\ref{e.magstring}) and the Hubbard one (\ref{e.maghub})
have the same exponential term. The aim of this paper is to understand also the
prefactor in (\ref{e.magstring}) which should be sensitive not only
to the kinematics but also to the details of the dynamics of the
worldsheet theory.

\section{Diagrammatic derivation of the prefactor}

In this section we will derive, using an adaptation of the
diagrammatic methods of \cite{LUSCHER,KM}, formulas for F- and
$\mu$-term corrections for a field theory with a generic dispersion
relation for elementary excitations. In general we have to make some
assumptions about 
normalizations of states and the form of Green's functions. These are
quite plausible but we do not prove them. In the course of the
calculation we will also assume that the same types of diagrams give
dominant finite size corrections as in the relativistic case. We thus
treat this derivation in a somewhat heuristic manner.

Our derivation follows closely the relativistic diagrammatic
derivation as discussed in \cite{KM}.

\subsubsection*{Green's function}
 
We assume that we are dealing with a 2D quantum field theory whose
elementary excitations follow a dispersion relation
\eq
E^2=\eps^2(p)
\eqx
In the relativistic case $\eps^2(p)=m^2+p^2$ while for the $AdS_5
\times S^5$ worldsheet theory $\eps^2(p)=1+8g^2 \sin^2 \f{p}{2}$. We note
the similarity between the two dispersion relations - both of which are
of the square-root type.

Let us consider a two point {\em Euclidean} Green's function for the
elementary excitations and its Fourier transform:
\eq
\cor{\phi_a(x) \phi_b(0)}=\dl_{ab} \int \f{d^2p}{(2\pi)^2} e^{ipx} G_a(p)
\eqx 
We define the self-energy $\Sg$ through
\eq
G_a(p)^{-1}=\eps_E^2+\eps^2(p)-\Sg
\eqx
where $\eps_E$ is the Euclidean energy. The Green's function should
have a pole along the mass shell manifold $\eps_E^2+\eps^2(p)=0$.
Moreover we fix its residue by analogy with relativistic case
\eq
\res_{\eps_E^2} G_a(p)=1  \qqqq \left(i.e.\;\;
\res_{\eps_E}=\f{1}{2\eps_E} \right)
\eqx
This means that both $\Sg$ and its partial derivatives vanish on
mass shell. In the following we will also need to know the appropriate
residue w.r.t. the spatial momentum which is
\eq
\label{e.resg}
\res_{p^1=\ps} G_a(p)=\f{1}{\eps^2(\ps)'}
\eqx 

Let us now put the theory on a cylinder of circumference $L$. Then the
self energy will get modified and one can compute the $L$-dependent
shift of the energy by examining the condition for the pole of the
Green's function
\eq
\eps_E^2+\eps^2(p)-\Sg_L(p)=0
\eqx
with $\eps_E=i(\eps(p)+\dl \eps_L)$. In this way we get the key
formula
\eq
\label{e.master}
\dl \eps_L = -\f{1}{2\eps(p)} \Sg_L(p)
\eqx 
We thus have to calculate the shift of the self energy due to finite
size effects.

\subsubsection*{Finite size self energy correction $\Sg_L(p)$}

When we put a theory on a cylinder of circumference $L$, the
coordinate space cylinder Green's function can be reconstructed from
averaging the infinite volume Green's function
over translations $x \to x+n L$. In momentum space one will get just
factors of $e^{inp^1 L}$ with $n \in \ZZ$ which should be redistributed
over all lines. We will now assume, following \cite{LUSCHER} in the
relativistic case, that the leading finite size correction arises from
a graph where only a {\em single} line has a factor
$e^{iq^1L}+e^{-iq^1L}$. Since we are integrating over the loop variable
$q$, we can substitute  $e^{iq^1L}+e^{-iq^1L} \to 2e^{-iq^1 L}$ under
the integral, for that single line. All
remaining parts of the graph are computed with infinite $L$ Feynman
rules.

\begin{figure}[t]
\centerline{\includegraphics[width=13cm]{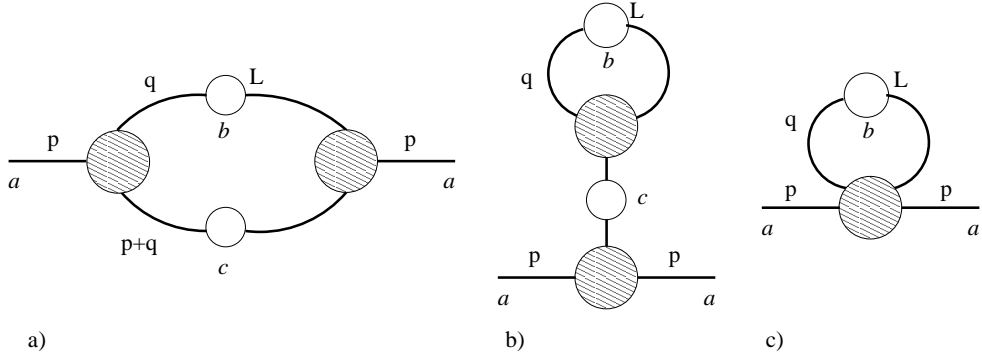}}
\caption{The graphs giving a leading finite size correction to the
  self energy: a) $I_{abc}$, b) $J_{abc}$, c) $K_{ab}$. The filled
  circles are the vertex functions $\Gamma$, empty circles represent the
  2-point Green's function. The letter $L$ represents the factor of
  $e^{-iq^1 L}$ and the letters in italics label the type of
  particles.}
\end{figure}

Following \cite{KM} we get three types of graphs contributing to the
self-energy of a particle of type $a$:
\eq
\Sg_L= \f{1}{2}\left( \sum_{bc} I_{abc} +\sum_{bc} J_{abc} +\sum_b
K_{ab} \right)
\eqx 
These graphs are shown in figure 2.
The corresponding expressions are
\eqn
I_{abc}\!\!\! &=&\!\!\! \int \f{d^2 q}{(2\pi)^2} 2e^{-iq^1 L} G_b(q)G_c(q+p)
\Gamma_{abc}(-p,-q,p+q) \Gamma_{acb}(p,-p-q,q) \nonumber\\
J_{abc} \!\!\!&=&\!\!\! \int \f{d^2 q}{(2\pi)^2} 2e^{-iq^1 L} G_b(q)
\Gamma_{bbc}(q,-q,0) G_c(0) \Gamma_{aac}(-p,p,0) \nonumber\\
K_{ab} \!\!\!&=&\!\!\! \int \f{d^2 q}{(2\pi)^2} 2e^{-iq^1 L} G_b(q)
\Gamma_{aabb}(p,-p,q,-q) 
\eqnx
where the $\Gamma$'s are the 3- and 4-point vertex functions.

The idea now is to shift the contour of integration over the {\em
  momentum} $q^1$ to imaginary values ($Im \; q^1=\kappa<0$
here). Then the contribution of the shifted contour will be
exponentially supressed by $e^{-\kappa L}$, which we will henceforth
  neglect. However, on the way, we 
get contributions from the poles of the propagators
$G_{.}(q)$. This forces the appropriate line to be {\em on-shell}
i.e. $q^1=\qs$ where
\eq
\label{e.mass}
{q^0_E}^2+\eps(\qs)^2=0
\eqx   
Let us work this out for the case of the $AdS_5 \times S^5$
superstring theory. Then using  $\eps(p)=\sqrt{1+8g^2 \sin^2 p/2}$ we
get
\eq
\qs =-i 2 \arcsinh \f{\sqrt{1+q^2}}{\sqg}
\eqx
where we denoted $q^0_E$ by $q$. Note that when we interpret the
  euclidean energy as momentum $p_{TBA}$, this is exactly $(\pm i)E_{TBA}$ that
  we obtain from the space-time interchange principle advocated in
  \cite{US}. Moreover the factor $e^{-iq^1 L}$ becomes
\eq
e^{-iq^1 L} \equiv e^{-i\qs L} = e^{-L \cdot E_{TBA}(q)}=e^{-L \cdot 2 \arcsinh
  \f{\sqrt{1+q^2}}{\sqg}}  
\eqx 
which coincides with the formula (\ref{e.heurf}) suggested in \cite{US}.

Before we proceed let us note that there is a subtlety associated with
the graph $I_{abc}$. There 
one can pick up two poles -- one associated with $G_b(q)$, the other
one associated with $G_c(q+p)$. We will denote the contribution of the
first pole by $I^+_{abc}$, and of the second one by
$I^-_{abc}$. 
It is convenient to shift the integration
variable in $I^-_{abc}$ as $q_{old}= q_{new}-p$. Then the value of the
momentum at the pole, $\qs$, will
be the same for all graphs $I^\pm_{abc}$, $J_{abc}$ and $K_{ab}$. 

The shift of the integration variable $q_{old}=
q_{new}-p$ has another important consequence. Since $q$ is Euclidean
while $p$ is Minkowskian, the contour of integration over Euclidean
energy got shifted into the complex plane.
During this process, one may encounter a pole. The additional
contribution of such a pole, evaluated by residues is exactly the
$\mu$-term. We will come back to this point at the end of this
section.

In addition, following \cite{KM}, let us change $I^-_{abc} \to I^-_{acb}$
since we are summing over $b$ and $c$ anyway. At this stage, after
taking the residue at (\ref{e.mass}) and using~(\ref{e.resg}) we are
left with
\eq
\Sg_L=\f{1}{2} \cdot 2 \cdot \int \f{dq^0_E}{2\pi} \f{i}{\eps^2(\qs)'}
\cdot e^{-|\qs|L} \cdot Integrand
\eqx
where the $Integrand$ is given by a sum of terms coming from
$I^+_{abc}+I^-_{acb}+ J_{abc}+K_{ab}$:
\eqn
Integrand &=& \sum_{bc} \Bigr( \Gamma_{abc}(-p,-q,p+q) G_c(p+q)
\Gamma_{acb}(p,-p-q,q)+ \nonumber\\
&+&  \Gamma_{acb}(-p,p-q,q) G_c(p-q) \Gamma_{abc}(p,-q,q-p) +
\nonumber\\
&+&  \Gamma_{aac}(p,-p,0) G_c(0) \Gamma_{bbc}(q,-q,0)+ \nonumber\\
&+&  \Gamma_{aabb}(p,-p,q,-q) \Bigl)
\eqnx
with {\em both} $p$ and $q$ being on-shell. Note that in the second
line, coming from $I^-_{acb}$, we made the appropriate shift of the
momentum $q$. 
The crucial observation is
now that the integrand is just an amputated connected 4-point forward
Green's function between on-shell particles (see e.g. \cite{KM}). We
thus get 
\eq
\label{e.sggreen}
\Sg_L= \int \f{dq^0_E}{2\pi} \f{i}{\eps^2(\qs)'} \cdot e^{-|\qs|L} \cdot
\sum_b G_{abab}(-p,-q,p,q) 
\eqx
In order to obtain the final expression it remains to connect the
forward Green's function with the on-shell S-matrix of the theory.

\subsubsection*{Link with the S-matrix}

The 4-point Green's function appearing in (\ref{e.sggreen}) is
essentially the S-matrix up to a
different normalization convention which we will now discuss.

We define the scalar product between asymptotic states in analogy to
the relativistic case as
\eq
\label{e.norm}
\cor{b(q)|a(p)}=\dl_{ab} 2 p^0 \cdot 2\pi \dl(p^1-q^1)
\eqx
The (forward) S-matrix appears as the scalar product between
2-particle {\em in} and {\em out} states, with the normalization
(\ref{e.norm}) factored out:
\eqn
\label{e.sdef}
{}_{out}\!\cor{b(q')a(p')|b(q)a(p)}_{in}&=& 4 p^0 q^0 (2\pi)^2
\dl({p^1}'-p^1) \dl({q^1}'-q^1)\, S_{ba}^{ba}(q,p) \nonumber \\
&\equiv & \dl_{fi} S_{ba}^{ba}(q,p)
\eqnx 
On the other hand the forward Green's function $G_{abab}(-p,-q,p,q)$
which is identified with the forward elastic amplitude
$T_{ab}(p,q|p,q)$ is defined through
\eq
\label{e.tdef}
{}_{out}\!\cor{b(q')a(p')|b(q)a(p)}_{in} =\dl_{fi}+i (2\pi)^2
\dl^{(2)}(p'+q'-p-q) T_{ab}(p,q|p,q)
\eqx 
where $\dl_{fi}$ is defined as in (\ref{e.sdef}). In order to express
$T$ through $S$ we have to calculate the Jacobian between the delta
functions appearing in (\ref{e.sdef}) and (\ref{e.tdef}):
\eq
\dl^{(2)}(p'+q'-p-q)=\f{1}{\eps'(p^1)-\eps'(q^1)} \cdot
\dl({p^1}'-p^1) \dl({q^1}'-q^1)  
\eqx
Putting these formulas together we get
\eq
\label{e.gands}
G_{abab}(-p,-q,p,q)=-4i \eps(p) \eps(\qs) (\eps'(\qs)-\eps'(p)) \left[
  S_{ba}^{ba}(q,p)-1 \right]
\eqx
which completes our derivation. For relativistic kinematics, the above
  formula reduces to the standard one
\eq
G_{abab}(-p,-q,p,q)=-4i m_a m_b \sinh(\th_q-\th_p) \left[
  S_{ba}^{ba}(\th_q-\th_p)-1 \right] 
\eqx

\subsubsection*{Final formulas}

Inserting the relation (\ref{e.gands}) into (\ref{e.sggreen}) and
(\ref{e.master}), we thus arrive at our final formula for the
generalized F-term finite size correction to the energy of a particle
$a$ with momentum $p$:   
\eq
\label{e.fgen}
\dl \eps^{F}_a = -\int_{-\infty}^\infty \f{dq}{2\pi}  \left(1-
\f{\eps'(p)}{\eps'(\qs)} \right) \cdot e^{-i\qs L} \cdot \sum_b
\left( S_{ba}^{ba}(\qs,p)-1 \right)
\eqx
Here $q$ is the original euclidean energy which plays the role of
momentum ($p_{TBA}$) in the space-time interchanged theory, $E=\eps(p)$ is the
dispersion relation and $\qs$ is determined by the on-shell condition
\eq
q^2+\eps^2(\qs)=0
\eqx

Let us now proceed to obtain the generalized form of the $\mu$-term
contribution. 
In the course of derivation of the formula (\ref{e.fgen}), we have moved the
contour of integration of the euclidean momentum into the complex
plane picking up a pole of the Green's function putting the particle
effectively on-shell and thus reducing the original double integral to
a {\em single} integral over euclidean energy which plays the role of
{\em momentum} in the space-time interchanged theory. 
However there was a subtlety that then in the
graph $I^-_{acb}$ the mass-shell condition was different from the one in
the remaining graphs. This
could be compensated by a shift of the $q$ contour. In doing so one
may encounter additional poles, the residues of which generate the
$\mu$-term contribution. Thus we obtain the generalized expression for
the $\mu$-term 
\eq
\label{e.mugen}
\dl \eps^{\mu}_a =-i \left(1-\f{\eps'(p)}{\eps'(\qts)} \right) \cdot
e^{-E_{TBA}(\qts) L} \cdot \res_{q=\qt} \sum_b S_{ba}^{ba}(\qs,p) 
\eqx  
where $\qt$ is the {\em Euclidean energy} of the pole of the S-matrix,
while $\qts$ is the corresponding momentum.

Before we apply the above formula to calculate leading finite size
corrections to the giant magnon dispersion relation in the following
section, let us illustrate how the generalized F-term formula
(\ref{e.fgen}) reduces to the ordinary one for the case of relativistic
kinematics i.e. when $\eps(p)=\sqrt{1+p^2}$. It is rather difficult to
keep track of all the choices of branches for the rapidity coordinates
when performing the Wick rotations and shifts of the contours so we
will not attempt to do it here. In addition there are various
conventions for the S-matrix (see e.g. \cite{DoreyMaldacena}). 
As in the case of the magnon
calculation presented in the remaining part of the paper we will
justify our choices {\em a-posteriori} by the final result. However it
would be very interesting to rigorously fix all such ambiguities from
first principles. We leave this extension for future work.
 
In the relativistic case one can reduce the formula (\ref{e.fgen}) to the
classical result in a simple manner when one retains a factor of
$e^{iq^1 L}$ and not $e^{-iq^1 L}$. Then the on-shell condition is
satisfied by 
\eqn
q\equiv -iq^0 &=&\sinh \th \equiv \cosh\left(\th+i\f{\pi}{2} \right)\\ 
\qs\equiv q^1&=&i\cosh\th \equiv \sinh\left(\th+i\f{\pi}{2} \right)
\eqnx 
In the relativistic case $\eps'(p)=p/\eps(p)=\tanh \th_p$. Thus we
obtain after some manipulations
\eq
\dl \eps^{F}_{relativistic}= \f{-1}{\cosh\th_p} \int_{-\infty}^\infty
\f{d\th}{2\pi} \cosh \left(\th-\th_p\right) \cdot e^{-L \cosh \th}
\cdot \sum_b \left( S_{ba}^{ba}\left(\th+i\f{\pi}{2}-\th_p \right) -1
\right) 
\eqx
which is exactly the formula derived in \cite{KM}. When the particle
is at rest, $\th_p=0$, we obtain the classical Luscher formula for
the finite size mass shift~(\ref{e.fluscher}).

\section{The giant magnon finite $J$ corrections}

Let us now apply the preceeding formalism to compute the leading finite
size corrections to the giant magnon. From the discussion in
section~3, we see that the exponential term of the classical correction is
captured by the $\mu$-term. We will now calculate the prefactor of the
exponential term using (\ref{e.mugen}). This requires taking the
residue at the BPS bound state \cite{Dorey} pole.

\subsection*{Kinematics}

In order to apply the generalized formula (\ref{e.mugen}) to the case
at hand we have to calculate the various kinematical factors appearing
in (\ref{e.mugen}) at the position of the BPS bound state pole.

The energy and momentum of a particle (magnon) in the worldsheet
theory of the superstring in $AdS_5 \times S^5$ is encoded in two
(complex) variables $x^\pm$ constrained by the equation
\eq
x^+ +\f{1}{x^+}-x^- -\f{1}{x^-} =i\f{\sqrt{2}}{g}
\eqx
The above equation defines a torus, so equivalently one may use a
parametrization by a single complex parameter (generalized rapidity)
in the complex plane being the universal covering space of the torus
\cite{CROSSING}. We will not, however, use this parametrization in the
present paper. 

The energy and momentum are then reconstructed from
\eq
e^{ip}=\f{x^+}{x^-} \qqqq E=\sqrt{2} g i(x^--x^+)-1
\eqx 
There exist inverse formulas expressing $x^\pm$ in terms of the
momentum, albeit with a branch cut ambiguity:
\eq
\label{e.xpmdef}
x^\pm=\f{1+\sqrt{1+8g^2 \sin^2 \f{p}{2}}}{\sqrt{8g^2} \sinpt} e^{\pm i
  \f{p}{2}}  
\eqx

We will need to find explicitly the $x_p^\pm$ parameters of the magnon
up to $\OO{1/g^2}$ accuracy. We obtain from (\ref{e.xpmdef}) 
\eqn
x_p^+ &=& e^{\f{ip}{2}} \left(1+\f{1}{\sqg \sinpt} +\OO{\f{1}{g^2}}
\right) \nonumber\\
x_p^- &=& e^{\f{-ip}{2}} \left(1+\f{1}{\sqg \sinpt} +\OO{\f{1}{g^2}}
\right)
\label{e.xp}
\eqnx
These expressions indeed satisfy the constraint equation up to $\OO{1/g^2}$
terms.
The corresponding BPS bound state is determined by the equation
$x_p^+=x_q^-$. Thus
\eq
\label{e.xqm}
x_q^-= e^{\f{ip}{2}} \left(1+\f{1}{\sqg \sinpt} +\OO{\f{1}{g^2}}
\right)
\eqx
and the corresponding $x_q^+$ can be found to be\footnote{Here we pick
a solution lying closest to the physical line.}
\eq
x_q^+= e^{\f{ip}{2}} \left(1+\f{3}{\sqg \sinpt} +\OO{\f{1}{g^2}}
\right)
\label{e.xqp}
\eqx
It will be useful to obtain explicit formulas for the momentum $\qts$
corresponding to the BPS pole as a function of $p$. By definition
\eq
e^{i \qts} \equiv \f{x^+_q}{x^-_q} \sim 1+\f{1}{\sqrt{2 g^2} \sinpt}
\eqx
so
\eq
\label{e.qsp}
\qts \sim \f{-i}{\sqrt{2} g \sinpt}
\eqx
Using the above expression we at once get the exponential term of the
magnon finite size corrections, justifying from a different point of
view the estimate (\ref{e.expmu}) of section 3 :
\eq
e^{-i \qts L} \sim e^{-\f{1}{\sqrt{2} g \sinpt} L} \equiv e^{-\f{2\pi
    J}{\sqrt{\lm} \sinpt}}
\eqx
where we identified $L$ with $J$ (we will come back to this point
when discussing general uniform light-cone gauges at the end of the paper).

We are now ready to evaluate the final nontrivial missing kinematical
factor in (\ref{e.mugen}) namely $\eps'(\qts)$. To this end we have
\eq
\eps'(\qts)=\f{4 g^2 \sin \f{\qts}{2} \cos \f{\qts}{2}}{\sqrt{1+8g^2
    \sin^2 \f{\qts}{2}}} \sim \f{\sqrt{2} g}{\cos \f{p}{2}} +\OO{1}
\eqx
Combining the above formula with the strong coupling expansion of the
derivative of $\eps(p)$ which can be calculated to be 
\eq
\eps'(p)=\f{4 g^2 \sin \f{p}{2} \cos \f{p}{2}}{\sqrt{1+8g^2 \sin^2 \f{p}{2}}}
\sim g \sqrt{2} \cos \f{p}{2} +\OO{\f{1}{g}}
\eqx
we obtain finally
\eq
1-\f{\eps'(p)}{\eps'(\qts)}  \sim 1-\cos^2 \f{p}{2}=\sin^2 \f{p}{2} 
\eqx
Putting together the above kinematical factors,
the resulting formula for the $\mu$-term becomes
\eq
\dl \eps^{\mu}_a =-i \cdot \sin^2 \f{p}{2} \cdot e^{-\f{1}{\sqrt{2} g
    \sinpt} L}  \cdot \res_{\qs=\qts} \sum_b S_{ba}^{ba}(\qs,p) 
\eqx
We will now proceed to evaluate the residue of the forward S matrix at
the BPS pole.

\subsection*{The S-matrix contribution}

We will concentrate on the finite size corrections to the magnon in
the {\sl su(2)} subsector. The S-matrix has the following form
\eq
S(x_q,x_p)=S_0(x_q,x_p) \cdot \left[ \hat{S}(x_q,x_p) \otimes
  \hat{S}(x_q,x_p) \right]
\eqx  
where $\hat{S}(x_q,x_p)$ is the $su(2|2)$ invariant S-matrix while the
scalar factor $S_0(x_q,x_p)$ is expressed in terms of the dressing factor
as
\eq
S_0(x_q,x_p)=\f{x_q^- - x_p^+}{x_q^+-x_p^-} \f{1-\f{1}{x_q^+ x_p^-}}{1-
  \f{1}{x_q^- x_p^+}} \cdot \sg^2(x_q,x_p)
\eqx
We will leave the evaluation of the contribution of the dressing
factor $\sg^2(x_q,x_p)$ to the next section, concentrating now on the
remaining matrix structure. 

The sum $\sum_b \left[\hat{S}\otimes \hat{S}\right] _{ba}^{ba}(\qs,p)$
carried out for $a$ in the {\sl su(2)} sector becomes
\eq
(2a_1+a_2+2a_6)^2
\eqx
where the $a_i$'s are the coefficients introduced in \cite{ZF}, which
parametrize the $su(2|2)$ S-matrix. The ones relevant for us
are
\eqn
a_1 &=& \f{x_p^--x_q^+}{x_p^+-x_q^-} \f{\eta_q \eta_p}{\etat_q \etat_p} \\ 
a_2 &=& \f{(x_q^--x_q^+)(x_p^--x_p^+)(x_p^--x_q^+)}{(x_q^--x_p^+) 
  (x_q^- x_p^- -x_q^+ x_p^+)}  \f{\eta_q \eta_p}{\etat_q \etat_p} \\ 
a_6 &=& \f{x_q^+-x_p^+}{x_q^--x_p^+} \cdot \f{\eta_p}{\etat_p}
\eqnx 
In the above expressions there are certain phase factors which depend
on the choice of basis for multiparticle states. It will be crucial
for us to use the choice corresponding to the {\em string frame} of
\cite{ZF} which is
\eqn
\f{\eta_q}{\etat_q} &=& e^{i \f{p}{2}} \equiv \sqrt{\f{x_p^+}{x_p^-}}
\\
\f{\eta_p}{\etat_p} &=& e^{-i \f{\qs}{2}} \equiv \sqrt{\f{x_q^-}{x_q^+}}
\eqnx 
Note that the phase factors are nonlocal -- it is this feature that
enables one to have an untwisted Yang-Baxter equation. In the {\em
  spin chain frame} corresponding to the original derivation of the
$su(2|2)$ S-matrix in \cite{Beisert} these factors are equal to unity.

We thus have to calculate the residue 
\eq
\res_{q=\qt}\, S_0(\qs,p) \cdot (2a_1+a_2+2a_6)^2
\eqx 
at the BPS bound state pole. It is most convenient to factor out
$(x_q^--x_p^+)$ in the denominator and use the de l'Hospital rule to
calculate
\eq
\lim_{q \to \qt}\, (q-\qt) \cdot \f{1}{x_q^--x_p^+} =\f{1}{{x_q^-}'}
\eqx
The derivative ${x_q^-}'$ can be evaluated to
\eq
{x_q^-}' \equiv \f{dx_q^-}{dq} = \f{i}{4} e^{-i \f{p}{2}} \left(
1-e^{ip} \right)^2
\eqx
at the BPS bound state pole. The derivation of this formula is
summarized in Appendix A.

One can now insert the formulas (\ref{e.xp}-\ref{e.xqp}) into the
remaining part of the expression to obtain, in the strong coupling
limit
\eq
\res_{q=\qt} \sum_b S_{ba}^{ba}(\qs,p) = \f{4 \sqrt{2} i}{g \cdot
  \sin^3 \f{p}{2}} \cdot e^{ip} \cdot \sg^2(x_q,x_p)
\eqx
The phase $e^{ip}$ is due to the `string frame` phase factors. It will
be crucial to cancel an analogous phase from the AFS part of the
dressing factor. If one would just use the BDS S-matrix for the {\sl
  su(2)} sector only, the result would be
\eq
\res_{q=\qt} S_{BDS}(\qs,p) = \f{ \sqrt{2} i}{g \cdot
  \sin^3 \f{p}{2}}
\eqx
The full expression for the $\mu$-term finite size correction is
\eq
\label{e.partial}
\dl \eps^{\mu}_a=\f{1}{g} \cdot \f{4\sqrt{2}}{\sin \f{p}{2}} e^{ip} \cdot
\sg^2(x_q,x_p) \cdot e^{-\f{1}{\sqrt{2} g \sinpt} L}
\eqx 
while in the BDS case we get
\eq
\dl \eps^{\mu}_{BDS}=\f{1}{g} \cdot \f{\sqrt{2}}{\sin \f{p}{2}} \cdot
e^{-\f{1}{\sqrt{2} g \sinpt} L} 
\eqx 
It is interesting to compare the two expressions with the results for
the finite size magnon corrections (\ref{e.magstring}) and Hubbard
model calculation (\ref{e.maghub}). In both cases we obtain the
correct exponential term which, as we saw, is expected to be very
generic. For the prefactors the situation is different. 
In the case of the magnon, ignoring the dressing phase, we
see that both the $g$ and $p$ dependence are different. Thus we may
expect crucial effects from the dressing factor which we will evaluate
in the following section. On the other hand both the $g$ and $p$
dependence match exactly with the result of Hubbard model. Only the
overall numerical coefficient is different. This is not completely
unexpected since we considered only the BDS magnon subsector of the
Hubbard model. It would be very interesting to perform a similar
computation with the full effective S-matrix and see whether one can
get an exact matching. We will not, however, consider this problem in
the current paper. 

In the following section we will complete the calculation of
(\ref{e.partial}) by evaluating the dressing factor $\sg^2(x_q,x_p)$
at the position of the BPS bound state pole.

\subsection*{The contribution of the dressing phase}

It remains to evaluate the dressing phase factor at the position of
the BPS bound state pole. It has the general structure\footnote{Where
  we took into account that $\chi(x_q^-,x_p^+)=0$ at the BPS pole.}
\eq
\sg^2(x_q,x_p)=e^{2i\left( \chi(x_q^-,x_p^-)-\chi(x_q^+,x_p^-)+
  \chi(x_q^+,x_p^+) \right)}
\eqx
where $\chi(x,y)=\tilde{\chi}(x,y)-\tilde{\chi}(y,x)$ is antisymmetric
and defined through the series expansion \cite{BHL,BES}
\eq
\chit(x,y)=\sum_{n=0}^\infty \f{1}{g_{BHL}^{n-1} } \chit^{(n)}(x,y)
\eqx  
where
\eq
g_{BHL}=\f{g}{\sqrt{2}}
\eqx
and
\eq
\label{e.defchin}
\chit^{(n)}(x,y) =\sum_{r=2}^\infty \sum_{s=r+1}^\infty
\f{-c^{(n)}_{r,s}}{(r-1) (s-1)} \f{1}{x^{r-1} y^{s-1}}
\eqx
with the coefficients $c^{(n)}_{r,s}$ given by the BHL/BES choice
\eq
c^{(n)}_{r,s}=\f{(1-(-1)^{r+s}) \zeta(n)}{2 (-2\pi)^n \Gamma(n-1)}
(r-1)(s-1) \f{\Gamma(\f{1}{2} (s+r+n-3)) \Gamma(\f{1}{2} (s-r+n-1))}{
\Gamma(\f{1}{2} (s+r-n+1)) \Gamma(\f{1}{2} (s-r-n+3))}
\eqx
for $n \geq 2$ and standard expressions for $n=0,1$.
The term with $n=0$ is the celebrated AFS phase \cite{AFS}, the
function $\chi^{(0)}(x,y)$ can be resummed to
\eq
\label{e.chiafs}
\chi^{(0)}(x,y)= -\f{g}{\sqrt{2}} \left(\f{1}{y}-\f{1}{x} \right)
\left(1-(1-xy) \log \left(1-\f{1}{xy}\right) \right)
\eqx 
while $n=1$ is the `1-loop' HL correction \cite{HL}. We will first discuss
these two terms separately and then calculate the contribution of
$\chi^{(n)}$ with $n \geq 2$.

\subsubsection*{AFS phase contribution}

We can now substitute the expressions for $x_q^\pm$ and $x_p^\pm$ into
(\ref{e.chiafs}) to obtain
\eq
\label{e.sgafs}
\sg^2_{AFS}(x_q,x_p)=-\f{g^2}{2} \cdot e^{-ip} \cdot \sin^4 \f{p}{2}
\eqx
Let us note several salient features of this result. Firstly, there is
a factor of $g^2$ which, when inserted into (\ref{e.partial}), gives
the correct $g$ scaling of the magnon result. Secondly, the momentum
dependence changes from Hubbard-like (\ref{e.maghub}) to the one of
the magnon (\ref{e.magstring}). Thirdly, there is an overall complex
phase, which is exactly canceled by the phase factor choice of the
`string frame' of \cite{AFZ} in the S-matrix. 

At this stage, both the $g$ and $p$ dependence are exactly as in
(\ref{e.magstring}). The only difference is an overall numerical
coefficient. We will find that the remaining part of the dressing
factor will only give a numerical coefficient.

\subsubsection*{HL phase contribution}

The HL part of the dressing factor can be also resummed into a closed
form expression in terms of dilogarithm functions. Unfortunately it is
quite complicated to use these formulas for the evaluation at the BPS
bound state pole as these expressions have various branch cuts. One
can fix the cuts by comparision with the result of a direct
calculation of the double sum defining the HL phase factor. The latter
calculation can in fact be done analytically in the strong coupling
limit and we present it in Appendix~B. The result for the contribution
of the Hernandez-Lopez phase is thus
\eq
\label{e.sghl}
\sg^2_{HL}(x_q,x_p)=\f{1}{2}
\eqx

Putting together (\ref{e.sgafs}) and (\ref{e.sghl}) we obtain at this stage
\eq
\label{e.resafshl}
\dl \eps^{\mu}_a=-g \cdot \sqrt{2} \, \sg^2_{n\geq 2}(x_q,x_p) \cdot
\sin^3 \f{p}{2} \cdot e^{-\f{1}{\sqrt{2} g \sinpt} L}
\eqx
while the string result (\ref{e.magstring}) is
\eq
\dl \eps^{\mu}_a=-g \cdot \sqrt{2} \, \f{8}{e^2} \cdot \sin^3 \f{p}{2} \cdot
e^{-\f{1}{\sqrt{2} g \sinpt} J}
\eqx
It remains to calculate the contribution of `higher-loop' terms in the
dressing factor $\sg^2_{n\geq 2}(x_q,x_p)$.

\subsubsection*{Contribution of $\chi^{(n)}$ with $n\geq 2$}

Naively it may seem that $\chi^{(n)}$ with $n\geq 2$ should not
contribute at strong coupling, however this turns out not to be the
case. At strong coupling $x_q^\pm$ and $x_p^\pm$ are just at the edge
of the radius of convergence of the appropriate power
series and approach a singularity which compensates the inverse power
of the coupling. Indeed, examining the resummed forms of $\chit^{(n)}(x,y)$
for $n<12$ we see that there is a common structure appearing in the
denominator:
\eq
\chit^{(n)}(x,y)=\f{\ldots}{g^{n-1} (1-xy)^{n-1}}
\eqx 
Therefore for $(x,y)=(x_q^-,x_p^-)$ and $(x,y)=(x_q^+,x_p^-)$ the
large inverse power of the coupling constant is canceled by the second
factor. We will thus obtain a nonvanishing result at strong coupling for
$\chi^{(n)}(x,y)$ with $x,y$ of the form
\eqn
x&=&e^{i\f{p}{2}} \left(1+\f{a}{g_{BHL} \sinpt} \right) \\
y&=&e^{-i\f{p}{2}} \left(1+\f{b}{g_{BHL} \sinpt} \right)
\eqnx

As shown in Appendix C, it turns out that the odd terms do not
contribute $\chi^{(2n+1)}(x,y)=0$ and the even ones give
\eq
\label{e.chitwon}
\chi^{(2n)}(x,y)=i (-1)^n \f{(2n-2)!}{2^{4n-2} \pi^{2n}} \zeta(2n)
\f{1}{(a+b)^{2n-1}} \equiv \chi^{(2n)}(a+b)
\eqx
The remaining part of the dressing factor contribution is therefore
\eq
\label{e.sgntwo}
\sg^2_{n\geq 2}=\exp \left(2i \sum_{n=1}^\infty \chi^{(2n)}
\left(\f{1}{2}\right) - \chi^{(2n)} (1) \right)
\eqx
The above series turns out to be only asymptotic and as it stands is
divergent. However it can be resummed using Borel resummation which in
this case amounts to performing the following procedure on the sums
appearing in~(\ref{e.sgntwo}) 
\eqn
\sum_{n=1}^\infty a_n &=&\sum_{n=1}^\infty \f{a_n}{(2n-2)!}\cdot (2n-2)!
=\sum_{n=1}^\infty \f{a_n}{(2n-2)!}\cdot \int_0^\infty t^{2n-2} e^{-t}
dt =\nonumber\\
&=& \int_0^\infty \left(\sum_{n=1}^\infty \f{a_n}{(2n-2)!}
t^{2n-2}\right) \cdot e^{-t} dt
\eqnx
A Borel resummed sum of the asymptotic series is then {\em defined} by the
last expression:
\eq
\sum_{Borel} \equiv \int_0^\infty \left( \sum_{n=1}^\infty \f{a_n}{(2n-2)!}
t^{2n-2}\right) \cdot e^{-t} dt 
\eqx
Performing the above procedure on the expression (\ref{e.sgntwo}) we
obtain
\eq
\sg^2_{n\geq 2}=\exp \left( \int_0^\infty \f{2-\f{t}{\sinh
    \f{t}{2}}}{t^2} e^{-t} dt \right) =\f{8}{e^2}
\eqx 
where we checked the last equality numerically using {\sl Maple} up to 200
digit accuracy. We also proved this result analytically and we give
the proof in Appendix D.

We thus obtain finally 
\eq
\sg^2_{n\geq 2}=\f{8}{e^2}
\eqx
which when inserted into (\ref{e.resafshl}) reproduces exactly the finite
size correction to the magnon dispersion relation at strong coupling
(\ref{e.magstring}). Let us note that the agreement involves a
contribution from all (even) loop orders in the BHL/BES dressing
factor and is thus a very nontrivial test of the proposed expressions
together with the Borel resummation procedure. It would be very
interesting to obtain the same result directly from the convergent
summation/integral formulas in \cite{BES,DK,DoreyMaldacena}.

\subsection*{General $a$-gauges}

Up till now we considered finite size correction to the magnon
dispersion relation evaluated in a gauge which is just one member ($a=0$)
of the family of generalized light cone gauges where the
density of
\eq
P_+=J+a(\Delta-J) \equiv J+a E
\eqx
is kept fixed on the worldsheet. Changing $a$, changes the behaviour of
elementary degrees of freedom. However such physical properties as the
spectrum of energies, {\em after} one incorporates the level matching
condition, should be independent of $a$. Despite that, it is also
interesting to see if one can also directly understand the
$a$-dependent properties of the elementary excitations. 

In \cite{AFZ} finite size corrections to the magnon dispersion
relation were evaluated in an arbitrary $a$-gauge with the result:
\eq
\label{e.agauge}
\dl \eps=-g \f{8\sqrt{2}}{e^2} \sin^3 \f{p}{2} \cdot
e^{-\f{1}{\sqrt{2} g \sinpt} J} \cdot e^{-a p \cot\f{p}{2}}
\eqx

Let us see how this result is reproduced from the formalism of the
present paper. The S-matrix in an arbitrary $a$-gauge is related to
the one in $a=0$ gauge, which we considered up to this point, by a
simple scalar factor:
\eq
S_{a\neq 0}(x_q,x_p) = e^{-ia(\eps_q p -\eps_p \qs)} \cdot S_{a=0}(x_q,x_p)
\eqx
We will now evaluate this scalar factor. In the strong coupling
limit $\eps_p \sim \sqrt{8g^2} \sinpt$, while $\eps_q$ can be easily
found from 
\eq
\label{e.epsqs}
\eps_q =\sqrt{2} g i (x^-_q-x_q^+)-1 \sim -i \cot \f{p}{2}
\eqx
The momentum $\qs$ is
\eq
\qs =-i \log \f{x^+_q}{x_q^-} \sim \f{-i}{\sqrt{2g^2} \sinpt}
\eqx
We thus see that from the change in the S-matrix we will get an
additional factor
\eq
e^{-a p\cot \f{p}{2}} \cdot e^{2a}
\eqx
The first factor is exactly the $a$ dependent correction to the magnon
dispersion relation. The second factor will cancel with another source
of $a$ dependence -- namely the fact that the length of the string is
no longer equal to $J$ but is rather $L=J+aE$. Therefore the
exponential term will get modified to
\eq
e^{-\f{1}{\sqrt{2} g \sinpt} L} \equiv e^{-\f{1}{\sqrt{2} g \sinpt}
  (J+\eps_p)} \sim e^{-\f{1}{\sqrt{2} g \sinpt} J} \cdot e^{-2a}
\eqx 
Putting the above terms together we recover exactly (\ref{e.agauge}).

\section{Conclusions}

The aim of the present paper was to derive a generalization of
L\"{u}scher formulas for finite size corrections in relativistic
quantum field theories which could be applied to the worldsheet
integrable theory of the $AdS_5\times S^5$ superstring. In \cite{US}
it was suggested that such corrections, coming from virtual particles
going around the worldsheet cylinder are the string counterparts of
gauge theoretical wrapping interactions which go beyond the asymptotic
Bethe ansatz. The string theory point of view is very fruitful here as
such effects and corrections to the Bethe ansatz are inherent to the very
structure of quantum field theory of which the worldsheet theory is an
example. This is in contrast to the spin chain language where there
seems to be no guiding principle for incorporating wrapping
interaction effects.

We have shown that generic exponential terms describe well the
magnitude of various types of finite size corrections both at weak and
at strong coupling. The new result here is the very universal origin
of the exponential term in the finite size correction to the magnon
dispersion relation which just depends on the infinite volume
dispersion relation.

We adapted the diagrammatic arguments of \cite{LUSCHER,KM} to derive
the generalized formulas for the F- and $\mu$-terms, recovering in the
course of derivation the space-time interchange appearing as the
motivation of the exponential terms proposed in \cite{US}.

Finally we used the formula for the generalized $\mu$-term to evaluate
finite size corrections to the giant magnon dispersion relation
derived in \cite{AFZ,Gordon}. We found that the dressing factor had a
crucial contribution and is responsible for the difference in the
finite size structure of the Hubbard and string result. Moreover we
found that we had to use the S-matrix in the so-called string frame
\cite{ZF} in order to cancel a complex phase from the AFS dressing
phase.

We found that all even loop-orders of the BHL/BES dressing factor
contribute to the resulting expression, which after Borel resummation
exactly reproduces the result of \cite{AFZ} coming from a completely
independent classical string calculation.

Finally we considered the same calculation in a general $a$-gauge,
again reproducing exactly the result of \cite{AFZ}.  

It would be very interesting to apply the formalism of the present
paper to other cases, as well as to generalize it to multiparticle
states. Ultimately, one would like to obtain exact results valid for
any $J$ in analogy to similar treatments for certain relativistic
integrable field theories.

\bigskip

\noindent{\bf Acknowledgments.} We thank Sergey
Frolov for discussions and suggestions to consider general $a$-gauges. This
work has been supported in part by Polish Ministry of Science and
Information Technologies grant 1P03B04029 (2005-2008), RTN network
ENRAGE MRTN-CT-2004-005616, and the Marie Curie ToK KraGeoMP (SPB
189/6.PRUE/2007/7).

\appendix

\section{Evaluation of ${x_q^-}'$}

We have to compute the derivative of $x_q^-$ w.r.t. the Euclidean
energy $q$. It turns out to be convenient to rewrite both quantities
as functions of $p$ and evaluate
\eq
{x_q^-}' =\f{\f{dx^-_q}{dp}}{\f{dq}{dp}}
\eqx
At strong coupling we may use formula (\ref{e.xqm}) and get
\eq
\f{dx^-_q}{dp} \sim \f{i}{2} e^{i\f{p}{2}}
\eqx
The Euclidean energy $q$ is related to $\eps(\qs)$ through
$q=i\eps(\qs)$, and the latter quantity has already been
evaluated in terms of $p$ in (\ref{e.epsqs}). Taking the derivative 
\eq
\f{dq}{dp}=- \f{1}{2\sin^2 \f{p}{2}}
\eqx
we obtain finally
\eq
{x_q^-}' =\f{dx^-_q}{dq}= -i e^{i \f{p}{2}} \sin^2 \f{p}{2} =\f{i}{4}
e^{-i\f{p}{2}} (1-e^{ip})^2
\eqx

\section{Evaluation of $\sg^2_{HL}(x_q,x_p)$}

In this section we want to derive
	\eq
	\sigma_{HL}^{2}(x_{p},x_{q})=e^{2i(\chi^{(1)}(x_{q}^{-},x_{p}^{-})
	-\chi^{(1)}(x_{q}^{+},x_{p}^{-})+\chi^{(1)}(x_{q}^{+},x_{p}^{+}))} 
	\eqx
 analytically. We start with the expression for $c^{(1)}_{r,s}$
	\eq
	c^{(1)}_{r,s}=-\frac{(1-(-1)^{r+s})}{\pi}\frac{(r-1)(s-1)}{(s+r-2)
	(s-r)}
	\eqx
which we substitute into (\ref{e.defchin}) and introduce a new
summation index $2k=s-r-1$ to obtain 
	\eq
\label{e.appb}
	\tilde{\chi}^{(1)}(x,y)=\frac{1}{\pi}\sum_{r=2}^{\infty}
	\sum_{k=0}^{\infty}\frac{1}{2k+1}\frac{1}{y^{2k+1}}
	\frac{z^{r-1}}{r+k-1}   
\eqx
where $z=\frac{1}{xy}$.

Firstly let us focus on $\chi^{(1)}(x_{q}^{+},x_{p}^{+})$. At strong
coupling we have $x_{q}^{+}=x_{p}^{+}=e^{\frac{ip}{2}}$ and 
	\eq
	\chi^{(1)}(x_{q}^{+},x_{p}^{+})=\tilde{\chi}^{(1)}(x_{q}^{+},
	x_{p}^{+})-\tilde{\chi}^{(1)}(x_{p}^{+},x_{q}^{+})=0 
	\eqx
For the two remaining terms $z$ is close to $1$ at strong
coupling. Indeed we have
	\eq
	z= 1-\frac{a+b}{g_{BHL}\sin\frac{p}{2}} 
	\eqx
where $a+b=1$ for $(x,y)=(x_{q}^{+},x_{p}^{-})$ and $a+b=\frac{1}{2}$
for $(x,y)=(x_{q}^{-},x_{p}^{-})$. Performing the summation over $r$
in (\ref{e.appb})	 
	\eq
	\tilde{\chi}^{(1)}(x,y)=\frac{1}{\pi}\sum_{k=0}^{\infty}
	\frac{1}{2k+1}\frac{1}{y^{2k+1}}
	\frac{_{2}F_{1}(1,k+1;k+2;z)}{k+1}z 
	\eqx
and we can use relations for hypergeometric functions to change
argument $z\rightarrow 1-z$ and expand the result around
$1-z=0$. Leaving only the leading term we have 
	\eq
	_{2}F_{1}(1,k+1;k+2;z)z\sim -(k+1)\log(1-z)
	\eqx
Using in addition the relation
	\eq
	\sum_{k=0}^{\infty}\frac{x^{2k+1}}{2k+1}=\arctanh x
	\eqx
we get the final form of the function $\tilde{\chi}^{(1)}(x,y)$
	\eq
	\tilde{\chi}^{(1)}(x,y)=\frac{-1}{\pi}\log(\frac{a+b}{g_{BHL}
	\sin\frac{p}{2}})\arctanh(\frac{1}{y})  
	\eqx
which gives after the antisymmetrization
	\eq
	\chi^{(1)}(x,y)=\frac{-1}{\pi}\log(\frac{a+b}{g_{BHL}
	\sin\frac{p}{2}})\big(
	\arctanh(\frac{1}{y})-\arctanh(\frac{1}{x}) \big) 
	\eqx

It is easy to notice that at the strong coupling we have
$y=\frac{1}{x}$ in the cases which we consider. Using the well-known
relation 
	\eq
	\arctanh w=\frac{1}{2}(\log(1+w)-\log(1-w))
	\eqx
we derive	
	\eq
	\chi^{(1)}(x,y)=\frac{-1}{\pi}\log(\frac{a+b}{g_{BHL}
	\sin(\frac{p}{2})})\frac{i\pi}{2} 
	\eqx
and finally
	\eq
	\chi^{(1)}(x_{q}^{-},x_{p}^{-})-\chi^{(1)}(x_{q}^{+},x_{p}^{-})
	=\frac{i}{2}\log(2)
	\eqx
This gives the HL part of the dressing factor contribution at the BPS pole
	\eq
	\sigma^{2}_{HL}=e^{2i\frac{i}{2}\log(2)}=\frac{1}{2}
	\eqx

\section{Evaluation of $\chi^{(n)}(x,y)$ for $n\geq2$}

In this section we want to derive $\chi^{(n)}(x,y)$ for $n\geq 2$ at
strong coupling. Let $x,y$ be defined by the relations (57) and (58) 
	\begin{eqnarray*}
	 x&=&e^{i p}(1+\frac{a}{g_{BHL}\sin{\frac{p}{2}}})\\
	y&=&e^{-i p}(1+\frac{b}{g_{BHL}\sin{\frac{p}{2}}})
	\end{eqnarray*}

 Let us start with the even terms. Substituting the expression for
 $c^{(2m)}_{r,s}$ into (49) and introducing a new summation index
 $2k=s-r-1$  we obtain 
	\begin{eqnarray*}
	&&\tilde{\chi}^{(2m)}(x,y)=\\
	&=&\frac{1}{g_{BHL}^{2m-1}}\sum_{r=2}^{\infty}
	\sum_{k=0}^{\infty}\frac{-2\zeta(2m)}{2(2\pi)^{2m}
	\Gamma(2m-1)}\frac{\Gamma(k+r+m-1)}{     
	\Gamma(k+r-m+1)}\frac{\Gamma(k+m)}{\Gamma(k-m+2)}\frac{1}{
	(xy)^{r-1}y^{2k+1}}\\   
	&=&\frac{-1}{g_{BHL}^{2m-1}}\frac{\zeta(2m)}{(2\pi)^{2m}
	\Gamma(2m-1)y}W(\frac{1}{xy};\frac{1}{y^{2}})  
	\end{eqnarray*}
where
	\eq
	W(z,w)=\sum_{k=0}^{\infty}\frac{\Gamma(k+m)}{\Gamma(k-m+2)}
	w^{k}\sum_{r=2}^{\infty}\frac{\Gamma(k+r+m-1)}{\Gamma(k+r-m+1)}z^{r-1} 
	\eqx
The second sum can be expressed as a hypergeometric function
	\begin{eqnarray*}
	\sum_{r=2}^{\infty}\frac{\Gamma(k+r+m-1)}{\Gamma(k+r-m+1)}
	z^{r-1}=\frac{\Gamma(k+m+1)}{\Gamma(k-m+3)}  
	\ _{2}F_{1}(1,k+m+1,k-m+3,z) 
	\end{eqnarray*}
At strong coupling, $z=\frac{1}{xy}$ is close to 1. Indeed we have  
	\begin{equation}\label{wzor_na_z}
	z= 1-\frac{a+b}{g_{BHL}\sin\frac{p}{2}} 
	\end{equation}
Using relations for hypergeometric functions we can change argument
$z\rightarrow 1-z$ and expand the result around $1-z=0$. Leaving only
the leading term we get 
	\begin{eqnarray*}
	_{2}F_{1}(1,k+m+1,k-m+3,z)
	\sim \frac{\Gamma(k-m+3)\Gamma(2m-1)}{\Gamma(k+m+1)}(1-z)^{1-2m} 
	\end{eqnarray*}
Thus the sum over $r$ is independent of $k$ {\em at strong coupling} and equals
	\eq
	\sum_{r=2}^{\infty}\frac{\Gamma(k+r+m-1)}{\Gamma(k+r-m+1)}z^{r-1}
	 \sim \frac{\Gamma(2m-1)}{(1-z)^{2m-1}} 
	\eqx
At this stage we have
\eq
W(z,w)=\frac{\Gamma(2m-1)}{(1-z)^{2m-1}}
\sum_{k=0}^{\infty}\frac{\Gamma(k+m)}{\Gamma(k-m+2)}w^{k} 
\eqx
Let us now focus on the remaining sum over $k$
	\begin{eqnarray*}
	\sum_{k=0}^{\infty}\frac{\Gamma(k+m)}{\Gamma(k-m+2)}w^{k}&=&
	\sum_{k=0}^{\infty}(k+m-1)\cdot \ldots \cdot (k-m+2)
	w^{k-m+1}w^{m-1}\\ 
	&=&w^{m-1} \sum_{k=0}^{\infty}\frac{d^{2m-2}}{dw^{2m-2}}
	w^{k+m-1}=w^{m-1} \frac{d^{2m-2}}{dw^{2m-2}}
	\frac{w^{m-1}}{1-w}\\ 
	&=&w^{m-1}  \frac{\Gamma(2m-1)}{(1-w)^{2m-1}}\\
	\end{eqnarray*}
Putting these results together and plugging in the relation
(\ref{wzor_na_z}), we have 
	\eq
	\tilde{\chi}^{(2m)}(x,y)=\frac{-\Gamma(2m-1)}{g_{BHL}^{2m-1}}
	\frac{\zeta(2m)}{(2\pi)^{2m}}\frac{1}{(\frac{a+b}{g_{BHL}\sin(
	\frac{p}{2})})^{2m-1}}  
	\frac{1}{(y-\frac{1}{y})^{2m-1}} 
	\eqx
The leading term of the  $y$ expansion at strong coupling is equal to
$e^{-i p}$ then 
	\eq
	y-\frac{1}{y}=e^{-i
	\frac{p}{2}}-e^{i\frac{p}{2}}=-2i \sin(\frac{p}{2}) 
	\eqx
Summing up
	\eq
	\tilde{\chi}^{(2m)}(x,y)=i
	\Gamma(2m-1)\frac{\zeta(2m)}{(i
	\pi)^{2m}}\frac{1}{(a+b)^{2m-1}}\frac{1}{2^{4m-1}} 
	\eqx
Analogously we can derive 
	\eq
	\tilde{\chi}^{(2m)}(y,x)=\frac{-\Gamma(2m-1)}{g_{BHL}^{2m-1}}\frac{
	\zeta(2m)}{(2\pi)^{2m}}\frac{1}{(\frac{a+b}{g_{BHL}\sin(
	\frac{p}{2})})^{2m-1}}  
	\frac{1}{(x-\frac{1}{x})^{2m-1}} 
	\eqx
where $x=e^{i p}$ in the leading approximation. Then
	\eq
	\tilde{\chi}^{(2m)}(y,x)=-i
	\Gamma(2m-1)\frac{\zeta(2m)}{(i
	\pi)^{2m}}\frac{1}{(a+b)^{2m-1}}\frac{1}{2^{4m-1}} 
	\eqx
and we can see that functions $\tilde{\chi}^{(2m)}$ are
antisymmetric. At the end we get the relation 
	\eq
	\chi^{(2m)}(y,x)=i \Gamma(2m-1)\frac{\zeta(2m)}{(i
	\pi)^{2m}}\frac{1}{(a+b)^{2m-1}}\frac{1}{2^{4m-2}} 
	\eqx
which is in perfect agreement with (\ref{e.chitwon}).

For the odd term analogical derivation can be done and the result is 
	\begin{eqnarray*}
	\tilde{\chi}^{(2m+1)}(x,y)=-\Gamma(2m)\frac{\zeta(2m)}{(2\pi)^{2m}}
	\frac{1}{(\frac{a+b}{\sin(\frac{p}{2})})^{2m}}
	\frac{1}{(y-\frac{1}{y})^{2m}}\\ 
	=-\Gamma(2m)\frac{\zeta(2m)}{(2\pi)^{2m}}\frac{1}{(\frac{a+b}{\sin(
	\frac{p}{2})})^{2m}}
	\frac{1}{(x-\frac{1}{x})^{2m}}=\tilde{\chi}^{(2m+1)}(y,x) 
	\end{eqnarray*}
thus the functions $\tilde{\chi}^{(2m+1)}$ are symmetric and the
contribution to the dressing factor from the odd terms vanishes 
	\eq
	\chi^{(2m+1)}(x,y)=0
	\eqx

\section{Borel resummation of $\sg^2_{n\geq 2}$}

In this appendix we will perform a Borel resummation of the asymptotic series
	\eq
	\sigma^{2}_{n\geq 2} = \exp\Big( 2i
	\sum_{n=1}^{\infty}(\chi^{(2n)}(\frac{1}{2})-\chi^{(2n)}(1))\Big) 
	\eqx 
where
	\eq
	\chi^{(2n)}(a+b)=i \Gamma(2n-1)\frac{\zeta(2n)}{(i
	\pi)^{2n}}\frac{1}{(a+b)^{2n-1}}\frac{1}{2^{4n-2}} 
	\eqx
Firstly let us assume that $a+b=\frac{1}{2}$. Then
	\begin{eqnarray*}
	\sum_{n=1}^{\infty}(i  \frac{(2n-2)!}{2^{2n-1}}
	\frac{\zeta(2n)}{(i \pi)^{2n}} )&=& i \Re\Big(
	\sum_{k=2}^{\infty}(  \frac{(k-2)!}{2^{k-1}}
	\frac{\zeta(k)}{(i \pi)^{k}})\Big) =\\ 
	&\overset{Borel}{=}&i \Re\Big( \sum_{k=2}^{\infty}(
	\int_{0}^{\infty}( e^{-t} \frac{t^{k-2}}{2^{k-1}}
	\frac{\zeta(k)}{(i\pi)^{k}})dt)\Big)
\end{eqnarray*}
Using the definition of the $\zeta$ function,
$\zeta(k)=\sum_{m=1}^\infty 1/m^k$ we can rewrite the last expression
as
\eq
i \Re\Big( \frac{-1}{2\pi^2} \int_{0}^{\infty}(e^{-t}
	(\sum_{m=1}^{\infty}\frac{1}{m(m+\frac{i t}{2\pi})}))
	dt\Big)=i\Re\Big(\frac{i}{\pi}
	\sum_{m=1}^{\infty}\frac{\Gamma(0,-2i m\pi)}{k}  \Big). 
\eqx
where $\Gamma(x,y)$ is the incomplete $\Gamma$ function.

Analogously for $a+b=1$ we get
	\eq
	\sum_{n=1}^{\infty}(i  \frac{(2n-2)!}{2^{4n-2}}
	\frac{\zeta(2n)}{(i \pi)^{2n}}
	)=i\Re\Big(\frac{i}{\pi}
	\sum_{m=1}^{\infty}\frac{\Gamma(0,-4i m\pi)}{k}  \Big). 
	\eqx
Then
	\begin{eqnarray*}
	 \sigma^{2}_{n\geq 2} &=& \exp\Big( \frac{2}{\pi}
	\Re\big(-i \sum_{m=1}^{\infty}\frac{\Gamma(0,-2i
	m\pi)-\Gamma(0,-4i m \pi)}{m}\big)\Big)\\ 
	&=&\exp\Big(\frac{2}{\pi}\Re\big(i
	\int_{1}^{2}\frac{\log(1-e^{2i \pi t})}{t}dt\big)\Big)=\\ 
	&=&\exp\Big(\frac{2}{\pi}\Re\big(
	i\int_{1}^{2}(\frac{1}{t}\log(-16 \sin(\pi t))dt-\pi +
	3\frac{ \pi}{2}\log(2)\big)\Big) 
	\end{eqnarray*}
where $\int_{1}^{2}(\frac{1}{t}\ln(-16 \sin(\pi t))dt$ is {\em real-valued}
and well-defined integral and thus does not contribute. In this way
we obtain the final result  
	\eq
	\sigma^{2}_{n\geq 2} =\exp(-2+3\log(2))=\frac{8}{e^2}
	\eqx


\begin{thebibliography}{99}

\bibitem{Minahan:2002ve}
J.~A.~Minahan and K.~Zarembo,
``The Bethe-ansatz for N = 4 super Yang-Mills,''
JHEP {\bf 0303}, 013 (2003), [hep-th/0212208].

\bibitem{Beisert:2003tq}
N.~Beisert, C.~Kristjansen and M.~Staudacher,
``The dilatation operator of N = 4 super Yang-Mills theory,''
Nucl.\ Phys.\ B {\bf 664} (2003) 131, [hep-th/0303060].

\bibitem{Beisert:2003yb}
  N.~Beisert and M.~Staudacher,
  ``The N = 4 SYM integrable super spin chain,''
  Nucl.\ Phys.\ B {\bf 670} (2003) 439, [hep-th/0307042].

\bibitem{kor1}
  A.~V.~Belitsky, S.~E.~Derkachov, G.~P.~Korchemsky and A.~N.~Manashov,
  ``Quantum integrability in (super) Yang-Mills theory on the light-cone,''
  Phys.\ Lett.\ B {\bf 594} (2004) 385
  [hep-th/0403085].


\bibitem{Bena:2003wd}
  I.~Bena, J.~Polchinski and R.~Roiban,
  ``Hidden symmetries of the AdS(5) x S**5 superstring,''
  Phys.\ Rev.\ D {\bf 69} (2004) 046002,
  [hep-th/0305116].

                                                           
\bibitem{Dolan:2003uh}
  L.~Dolan, C.~R.~Nappi and E.~Witten,
  ``A relation between approaches to integrability in superconformal
  Yang-Mills theory,''
  JHEP {\bf 0310} (2003) 017, [hep-th/0308089].


\bibitem{Maldacena:1997re}
J.~M.~Maldacena,
``The large N limit of superconformal field theories and supergravity,''
Adv.\ Theor.\ Math.\ Phys.\  {\bf 2} (1998) 231
[Int.\ J.\ Theor.\ Phys.\  {\bf 38} (1999) 1113], [hep-th/9711200];\\
S.~S.~Gubser, I.~R.~Klebanov and A.~M.~Polyakov,
``Gauge theory correlators from non-critical string theory,''
Phys.\ Lett.\ B {\bf 428} (1998) 105, [hep-th/9802109];\\
E.~Witten,
``Anti-de Sitter space and holography,''
Adv.\ Theor.\ Math.\ Phys.\  {\bf 2} (1998) 253, [hep-th/9802150].


\bibitem{S}
  M.~Staudacher,
  ``The factorized S-matrix of CFT/AdS,''
  JHEP {\bf 0505}, 054 (2005)
  [hep-th/0412188].

\bibitem{BS}
  N.~Beisert and M.~Staudacher,
  ``Long-range PSU(2,2$|$4) Bethe ansaetze for gauge theory and strings,''
  Nucl.\ Phys.\ B {\bf 727}, 1 (2005)
  [hep-th/0504190].


\bibitem{Beisert}
  N.~Beisert,
  ``The su(2|2) dynamic S-matrix,''
  arXiv:hep-th/0511082.


\bibitem{AFS}
  G.~Arutyunov, S.~Frolov and M.~Staudacher,
  ``Bethe ansatz for quantum strings,''
  JHEP {\bf 0410}, 016 (2004)
  [hep-th/0406256].

\bibitem{HL}
  R.~Hernandez and E.~Lopez,
  ``Quantum corrections to the string Bethe ansatz,''
  JHEP {\bf 0607}, 004 (2006)
  [arXiv:hep-th/0603204].

\bibitem{BES}
  N.~Beisert, B.~Eden and M.~Staudacher,
  ``Transcendentality and crossing,''
  J.\ Stat.\ Mech.\  {\bf 0701}, P021 (2007)
  [arXiv:hep-th/0610251].

\bibitem{CROSSING}
  R.~A.~Janik,
  ``The AdS(5) x S**5 superstring worldsheet S-matrix and crossing symmetry,''
  Phys.\ Rev.\  D {\bf 73}, 086006 (2006)
  [arXiv:hep-th/0603038].

\bibitem{BHL}
  N.~Beisert, R.~Hernandez and E.~Lopez,
  ``A crossing-symmetric phase for AdS(5) x S**5 strings,''
  JHEP {\bf 0611}, 070 (2006)
  [arXiv:hep-th/0609044].

\bibitem{Lipatov}
  A.~V.~Kotikov and L.~N.~Lipatov,
  ``DGLAP and BFKL equations in the N = 4 supersymmetric gauge theory,''
  Nucl.\ Phys.\  B {\bf 661} (2003) 19
  [Erratum-ibid.\  B {\bf 685} (2004) 405]
  [arXiv:hep-ph/0208220].

\bibitem{ES}
  B.~Eden and M.~Staudacher,
  ``Integrability and transcendentality,''
  J.\ Stat.\ Mech.\  {\bf 0611}, P014 (2006)
  [arXiv:hep-th/0603157].

\bibitem{Bern}
  Z.~Bern, M.~Czakon, L.~J.~Dixon, D.~A.~Kosower and V.~A.~Smirnov,
  ``The Four-Loop Planar Amplitude and Cusp Anomalous Dimension in Maximally
  Supersymmetric Yang-Mills Theory,''
  Phys.\ Rev.\  D {\bf 75}, 085010 (2007)
  [arXiv:hep-th/0610248].

\bibitem{Zarembotwoloop}
  T.~Klose, T.~McLoughlin, J.~A.~Minahan and K.~Zarembo,
  ``World-sheet scattering in AdS(5) x S**5 at two loops,''
  arXiv:0704.3891 [hep-th].

\bibitem{DoreyMaldacena}
  N.~Dorey, D.~M.~Hofman and J.~Maldacena,
  ``On the singularities of the magnon S-matrix,''
  Phys.\ Rev.\  D {\bf 76}, 025011 (2007)
  [arXiv:hep-th/0703104].

\bibitem{BDS}
  N.~Beisert, V.~Dippel and M.~Staudacher,
  ``A novel long range spin chain and planar N = 4 super Yang-Mills,''
  JHEP {\bf 0407}, 075 (2004)
  [hep-th/0405001].

\bibitem{LSRV}
  A.~V.~Kotikov, L.~N.~Lipatov, A.~Rej, M.~Staudacher and V.~N.~Velizhanin,
  ``Dressing and Wrapping,''
  arXiv:0704.3586 [hep-th].

\bibitem{SN}
  S.~Schafer-Nameki,
  ``Exact expressions for quantum corrections to spinning strings,''
  Phys.\ Lett.\  B {\bf 639} (2006) 571
  [arXiv:hep-th/0602214].

\bibitem{SNZZ}
  S.~Schafer-Nameki, M.~Zamaklar and K.~Zarembo,
  ``How accurate is the quantum string Bethe ansatz?,''
  JHEP {\bf 0612}, 020 (2006)
  [arXiv:hep-th/0610250].

\bibitem{AFZ}
  G.~Arutyunov, S.~Frolov and M.~Zamaklar,
  ``Finite-size effects from giant magnons,''
  Nucl.\ Phys.\  B {\bf 778}, 1 (2007)
  [arXiv:hep-th/0606126].

\bibitem{US}
  J.~Ambjorn, R.~A.~Janik and C.~Kristjansen,
  ``Wrapping interactions and a new source of corrections to the spin-chain /
  string duality,''
  Nucl.\ Phys.\ B {\bf 736} (2006) 288
  [hep-th/0510171].

\bibitem{LUSCHER}
  M.~L\"{u}scher,
  ``Volume Dependence Of The 
Energy Spectrum In Massive Quantum Field Theories.
  1. Stable Particle States,''
  Commun.\ Math.\ Phys.\  {\bf 104} (1986) 177.

\bibitem{KM}
  T.~R.~Klassen and E.~Melzer,
  ``On The Relation Between Scattering Amplitudes And Finite Size Mass
  Corrections In Qft,''
  Nucl.\ Phys.\ B {\bf 362} (1991) 329.

\bibitem{magnon}
  D.~M.~Hofman and J.~M.~Maldacena,
  ``Giant magnons,''
  J.\ Phys.\ A  {\bf 39}, 13095 (2006)
  [arXiv:hep-th/0604135].

\bibitem{mannpol} 
  N.~Mann and J.~Polchinski,
  ``Bethe ansatz for a quantum supercoset sigma model,''
  Phys.\ Rev.\  D {\bf 72}, 086002 (2005)
  [arXiv:hep-th/0508232].

\bibitem{kazakov}
  N.~Gromov, V.~Kazakov, K.~Sakai and P.~Vieira,
  ``Strings as multi-particle states of quantum sigma-models,''
  Nucl.\ Phys.\  B {\bf 764}, 15 (2007)
  [arXiv:hep-th/0603043].

\bibitem{gromov}
  N.~Gromov and V.~Kazakov,
  ``Asymptotic Bethe ansatz from string sigma model on S**3 x R,''
  arXiv:hep-th/0605026.

\bibitem{srz}
  A.~Rej, M.~Staudacher and S.~Zieme,
  ``Nesting and dressing,''
  arXiv:hep-th/0702151.

\bibitem{sakai}
  K.~Sakai and Y.~Satoh,
  ``Origin of dressing phase in N=4 Super Yang-Mills,''
  arXiv:hep-th/0703177.


\bibitem{Gordon}
  D.~Astolfi, V.~Forini, G.~Grignani and G.~W.~Semenoff,
  ``Gauge invariant finite size spectrum of the giant magnon,''
  Phys.\ Lett.\  B {\bf 651}, 329 (2007)
  [arXiv:hep-th/0702043].

\bibitem{hubbard}
  A.~Rej, D.~Serban and M.~Staudacher,
  ``Planar N = 4 gauge theory and the Hubbard model,''
  [hep-th/0512077].

\bibitem{TBAGROUND}
  A.~B.~Zamolodchikov,
  ``Thermodynamic Bethe Ansatz In Relativistic Models. Scaling Three State
  Potts And Lee-Yang Models,''
  Nucl.\ Phys.\ B {\bf 342}, 695 (1990).

\bibitem{Dorey}
  P.~Dorey and R.~Tateo,
  ``Excited states by analytic continuation of TBA equations,''
  Nucl.\ Phys.\ B {\bf 482} (1996) 639, hep-th/9607167.

\bibitem{NLIE}
  C.~Destri and H.~J.~De Vega,
  ``Unified approach to thermodynamic Bethe Ansatz and finite size corrections
  for lattice models and field theories,''
  Nucl.\ Phys.\ B {\bf 438}, 413 (1995), hep-th/9407117.

\bibitem{fioravanti}
  D.~Fioravanti, A.~Mariottini, E.~Quattrini and F.~Ravanini,
  ``Excited state Destri-De Vega equation for sine-Gordon and restricted
  sine-Gordon models,''
  Phys.\ Lett.\  B {\bf 390}, 243 (1997)
  [arXiv:hep-th/9608091].

\bibitem{BLZZ}
  V.~V.~Bazhanov, S.~L.~Lukyanov and A.~B.~Zamolodchikov,
  ``Quantum field theories in finite volume: Excited state energies,''
  Nucl.\ Phys.\ B {\bf 489} (1997) 487, hep-th/9607099.


\bibitem{Teschner}
  J.~Teschner,
  ``On the spectrum of the Sinh-Gordon model in finite volume,''
  arXiv:hep-th/0702214.

\bibitem{ZF}
  G.~Arutyunov, S.~Frolov and M.~Zamaklar,
  ``The Zamolodchikov-Faddeev algebra for AdS(5) x S**5 superstring,''
  JHEP {\bf 0704}, 002 (2007)
  [arXiv:hep-th/0612229].

\bibitem{DK}
  I.~Kostov, D.~Serban and D.~Volin,
  ``Strong coupling limit of Bethe ansatz equations,''
  arXiv:hep-th/0703031.




\end{thebibliography}
\end{document}